\documentstyle[12pt,epsfig]{article}

\def\fig#1{Fig.~{\ref{#1}}}
\def\eqn#1{Eq.~(\ref{#1})}
\def\tree{{\rm tree}}
\newskip\humongous \humongous=0pt plus 1000pt minus 1000pt
\def\caja{\mathsurround=0pt}
\def\eqalign#1{\,\vcenter{\openup1\jot \caja
        \ialign{\strut \hfil$\displaystyle{##}$&$
        \displaystyle{{}##}$\hfil\crcr#1\crcr}}\,}
\newif\ifdtup

\newcounter{eqnumber}[section]
\renewcommand{\theeqnumber}{\thesection.\arabic{eqnumber}}
\def\equn{
\refstepcounter{eqnumber}
\eqno({\rm \theeqnumber})
}

\def\eqn#1{eq.~(\ref{#1})}

\def\fig#1{fig.~{\ref{#1}}}

\def\sec#1{section~{\ref{#1}}}

\def\app#1{appendix~\ref{#1}}

\def\tr{{\rm tr}}
\def\bra#1{\langle #1 |}
\def\ket#1{| #1 \rangle}
\def\bkmp#1{\langle #1 \rangle}
\def\bkpm#1{[ #1 ]}
\def\braket#1#2{\langle #1 |\  #2 \rangle}
\newbox\charbox
\newbox\slabox
\def\s#1{{      
        \setbox\charbox=\hbox{$#1$}
        \setbox\slabox=\hbox{$/$}
        \dimen\charbox=\ht\slabox
        \advance\dimen\charbox by -\dp\slabox
        \advance\dimen\charbox by -\ht\charbox
        \advance\dimen\charbox by \dp\charbox
        \divide\dimen\charbox by 2
        \raise-\dimen\charbox\hbox to \wd\charbox{\hss/\hss}
        \llap{$#1$}
}}
\def\A{{\cal A}}

\def\D{{\cal D}}
\def\N{{\cal N}}

\def\half{{1\over 2}}
\def\pol{\varepsilon}
\def\eps{\epsilon}

\def\Ord{{\cal O}}

\def\lsl{\s \ell}

\def\si{\sigma}

\def\spa#1.#2{\left\langle#1\,#2\right\rangle}
\def\spb#1.#2{\left[#1\,#2\right]}
\def\lor#1.#2{\left(#1\,#2\right)}
\def\sand#1.#2.#3{%
  \left\langle\smash{#1}{\vphantom1}\right|{#2}%
  \left|\smash{#3}{\vphantom1}\right\rangle}
\def\sandp#1.#2.#3{%
  \left\langle\smash{#1}{\vphantom1}^{-}\right|{#2}%
  \left|\smash{#3}{\vphantom1}^{+}\right\rangle}
\def\sandpp#1.#2.#3{%
  \left\langle\smash{#1}{\vphantom1}^{+}\right|{#2}%
  \left|\smash{#3}{\vphantom1}^{+}\right\rangle}
\def\sandmm#1.#2.#3{%
  \left\langle\smash{#1}{\vphantom1}^{-}\right|{#2}%
  \left|\smash{#3}{\vphantom1}^{-}\right\rangle}
\def\sandpm#1.#2.#3{%
  \left\langle\smash{#1}{\vphantom1}^{+}\right|{#2}%
  \left|\smash{#3}{\vphantom1}^{-}\right\rangle}
\def\sandmp#1.#2.#3{%
  \left\langle\smash{#1}{\vphantom1}^{-}\right|{#2}%
  \left|\smash{#3}{\vphantom1}^{+}\right\rangle}
\catcode`@=11  

\begin{document}

\begin{titlepage}

\begin{flushright}
hep-ph/9709423 \hfill UCLA/97/TEP/23\\
\today{}\\
\end{flushright}

\vskip 2.cm

\begin{center}
{\Large\bf Feynman Diagrams and Cutting Rules}
\vskip 2.cm

{\large J.S. Rozowsky\footnote{Email address: 
{\tt rozowsky@physics.ucla.edu}}}

\vskip 0.5cm

{\it  Department of Physics, University of California at Los
Angeles, CA 90095-1547}
\vskip 3cm
\end{center}

\begin{abstract}
We show how Feynman diagrams may be evaluated to take advantage of
recent developments in the application of Cutkosky rules to the
calculation of one-loop amplitudes.  A sample calculation of $gg
\rightarrow g H$, previously calculated by Ellis et~al., is presented
illustrating this equivalence. This example demonstrates the use of
cutting rules for massive amplitudes.
\end{abstract}

\vfill
\end{titlepage}

\section{Introduction}

One-loop calculations are important in the quest for discovering new
physics beyond the Standard Model, especially for determining QCD
backgrounds.  A number of significant improvements in the calculation
of one-loop amplitudes, especially ones with massless particles, have
been made in recent years.  These methods, which have recently been
reviewed in ref.~\cite{Review}, include the use of helicity
\cite{SpinorHelicity}, color decompositions \cite{Color},
string-inspired ideas \cite{Long}, recursion relations
\cite{Recursive,Mahlon}, unitarity
\cite{Cutting,SusyFour,SusyOne,Massive} and factorization
\cite{Factorization}.  Besides allowing for the computation of certain
infinite sequences of amplitudes, these techniques have been used in
the computation of non-trivial amplitudes relevant for
phenomenological applications \cite{FiveParton,Zjets}.

A technique that has proven to be especially useful is that of using
unitarity to determine the functional form of the amplitudes.  By
working to all orders in the dimensional regularization parameter,
$\eps= (4 -D)/2$, one can determine the coefficients of all integral
functions that may appear \cite{TwoLoopUnitarity}.  For certain
amplitudes that satisfy a power counting criterion, such as
supersymmetric ones, one can show that cuts with four-dimensional
momenta are sufficient to determine the amplitudes through
$\Ord(\eps^0)$ \cite{SusyFour,SusyOne}.  The case of an amplitude with 
massive internal legs has been considered in ref.~\cite{Massive}.

The efficiency of the unitarity method stems from the use of gauge
invariant tree amplitudes in performing calculations. With the use of
helicity and color decompositions this can lead to enormous reductions
in the complexity of intermediate expressions.  This may be contrasted
against calculations of individual Feynman diagrams that are gauge
variant, where large cancelations must take place in order obtain
simpler gauge invariant results.

In this paper we will show how to evaluate Feynman diagrams so as to
obtain efficiency of computation inherent in the unitarity technique.
This is done by performing Feynman diagram calculations in a
particular way so as to mimic the unitarity approach advocated in
ref.~\cite{Review}.  One may therefore start with Feynman diagrams,
yet obtain the advantages of the Cutkosky approach.  This is useful
since it explains the unitarity method from a conventional Feynman
diagram viewpoint.  This is of interest in, for example, the extension
of unitarity methods beyond one loop, such as in
ref.~\cite{BernYanRoz}.

As an explicit example we have calculated the one-loop helicity
amplitudes for $gg \rightarrow g H$.  This example has the desirable
features that it contains both colored and colorless lines and massive
internal and external legs.  This example is useful since it
illustrates features beyond purely massless calculations, where most
attention has been concentrated \cite{Review}.  Also it has already
been evaluated via conventional methods by R.K. Ellis et~al.
\cite{Ellis}, allowing for an easy comparison to the correct result to
be made.  We find agreement with their result.

\section{Review of basic tools}
\label{BasicTools}
We now briefly review some of the techniques that we shall use in this
paper. Two important techniques that we utilize are spinor helicity
and color decomposition. The reader may consult review articles
\cite{ManganoReview} for further details.

\subsection{Spinor helicity}
\label{SpinorHelicity}

In explicit calculations with external gluons, it is usually
convenient to use a spinor helicity basis \cite{SpinorHelicity} which
rewrites all polarization vectors in terms of massless Weyl spinors
$\vert k^{\pm} \rangle$.  In the formulation of Xu, Zhang and Chang
the plus and minus helicity polarization vectors are expressed as
$$
\pol^{+}_\mu (k;q) =  {\sandmm{q}.{\gamma_\mu}.k
      \over \sqrt2 \spa{q}.k}\, ,\hskip 1cm
\pol^{-}_\mu (k;q) =  {\sandpp{q}.{\gamma_\mu}.k
      \over \sqrt{2} \spb{k}.q} \, ,
\equn
\label{PolarizationVector}
$$  
where $k$ is the momentum of the gluon and $q$ is an arbitrary null
`reference momentum' which drops out of the final gauge invariant
amplitudes.  We use the convenient notation
$$
\langle k_i^{-} \vert k_j^{+} \rangle \equiv \langle ij \rangle \, , 
\hskip 2 cm 
\langle k_i^{+} \vert k_j^{-} \rangle \equiv [ij] \, .
\equn
$$ 
These spinor products are anti-symmetric and satisfy
$ \spa{i}.j \spb{j}.i = 2 k_i \cdot k_j$.
One useful identity is
$$
\sand{a}.{\s{\pol}^\pm(k;q)}.{b} =
\pm \frac{\sqrt{2}}{\braket{q^\mp}{k^\pm}}
\bra{a}  \left[
    \ket{q^\pm} \bra{k^\pm} + \ket{k^\mp} \bra{q^\mp}
  \right] \ket{b} \, ,
\equn\label{PolSlash}
$$
where either $\bra{a}$ or $\ket{b}$ are spinors with four-dimensional 
momenta.

It is convenient to choose a dimensional regularization scheme that
is compatible with the spinor helicity formalism.  We use the
four-dimensional helicity scheme \cite{Long} which is equivalent to a
helicity form of Siegel's dimensional reduction scheme \cite{Siegel}.
The conversion between the various dimensional regularization schemes
has been given in ref.~\cite{Long,KunsztFourPoint}.

\subsection{Color decomposition}
\label{ColorDecomposition}

Color decompositions \cite{Color} have been extensively discussed in
review articles \cite{ManganoReview}.  Here we only present the color
decomposition for the $gg \rightarrow g H$ amplitude.

For this one-loop amplitude 
the decomposition is%
$$
\A_4^{\rm 1-loop} =  g^3\mu_R^{2\eps} \sum_{\sigma}
\tr[T^{a_{\si(1)}}T^{a_{\si(2)}}T^{a_{\si(3)}}] \,
A^f_4 (\si(1), \si(2), \si(3), H) \, , 
\equn\label{Decomp}
$$
where we have taken the loop to be in the fundamental representation.
The sum over $\sigma$ includes all cyclic permutations of the indices
$\sigma(n)$ (i.e. $\sigma \in \{(1,2,3),(1,3,2)\}$) and the $T^a$ are
fundamental representation color matrices (normalized so that $\tr(T^a
T^b) = \delta^{ab}$).  We have abbreviated the dependence of $A^f_{4}$
on the outgoing momenta $k_j$ and helicities $\lambda_j$ by writing
the label $j$ alone.  We have also explicitly extracted the coupling
and a factor of $\mu_R^{2\eps}$ from $A^f_4$, where $\mu_R$ is the
renormalization scale.  For adjoint representation loops there is an
analogous decomposition with up to two color traces in each term.

It is convenient to further decompose the partial amplitude in terms
of {\it primitive amplitudes}, 
$$
\eqalign{
& A^f_{4} (\si(1), \si(2), \si(3), H)  \cr 
& \hskip 1.5 cm = A_{4} (\si(1), \si(2), \si(3), H) +
A_{4} (\si(3), \si(1), \si(2), H) +A_{4} (\si(2), \si(3), \si(1), H)
\,.   }
\equn\label{partialdecomposition}
$$
In the primitive amplitudes appearing on the right-hand-side we may
treat the Higgs as carrying color charge (in the adjoint
representation), which couples to gluons; it is not difficult to show
in the permutation sum all such additional terms cancel.  This is
convenient since these primitive amplitudes are gauge invariant and
have a fixed cyclic ordering of external legs.  A similar
decomposition in terms of `primitive amplitudes' can also be performed
for the case where some of the external particles are in the
fundamental representation \cite{Fermion}.

\section{Feynman diagrams}
\label{Feynman}

Feynman diagrams, which have been essential for evaluating terms in
the perturbative series for scattering processes, suffer from a number
of well known computational difficulties. Firstly, individual Feynman
diagrams are inherently gauge dependent; gauge invariance is restored
only after summing over all diagrams.  Secondly, when evaluating loop
momentum integrals using traditional techniques, such as
Passarino-Veltman reduction \cite{PV}, one encounters a large number
of spurious denominators. These spurious denominators typically
involve Gram determinants, $\det(k_i \cdot k_j)$ raised to a power.
Enormous cancelations must then take place before one obtains
relatively simple results.

When using conventional techniques to evaluate Feynman diagrams one
reduces all tensor integrals into sums of scalar integrals multiplied
by reduction coefficients.  (By a tensor integral we mean loop
momentum integrals with powers of momentum in the numerator and by a
scalar integral we mean integrals with no powers of loop momentum in
the numerator, see \app{IntegralsAppendix}.)  With the knowledge of
how this reduction works, given any one-loop amplitude we can list all
the scalar integrals that could potentially enter.  Thus we have
$$
\mbox{Amplitude} = \sum \mbox{Feynman diagrams} \rightarrow 
\sum_i c_i I_i\, , 
\equn\label{temp1}
$$
where the subscript $i$ is some index that labels the scalar integrals
that contribute to this amplitude. The heart of the calculation is in
evaluating all the coefficients of the scalar integrals, i.e. the
$c_i$'s.

A conventional Feynman diagram technique like Passarino-Veltman (PV)
reduction evaluates all the necessary coefficients simultaneously.  If
one could obtain these coefficients in a more efficient manner one
would greatly improve the ability to calculate.

Some of the features that are useful in improved calculational methods
are: (1) that the building blocks be gauge invariant combinations of
diagrams instead of individual diagrams; (2) that wherever possible
tensor integrals should not generate large powers of Gram determinants
in denominators; (3) that large cancelations between diagrams be
avoided as much as possible; (4) and that calculations can be recycled
when performing new ones.

\subsection{Review of the unitarity method}

\begin{figure}
\begin{center}
\epsfig{file=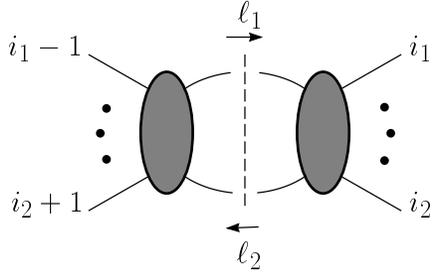,clip=,width=2.3in}
\end{center}
\vskip -1.75 cm
\caption[]{
\label{CutFigure}
A cut of a general $n$-point one-loop amplitude.
}
\end{figure}

The cutting method reviewed in ref.~\cite{Review} has these above
properties.  In this method one writes down the cuts in a given
channel, depicted in \fig{CutFigure},
$$
\eqalign{
A_{n}&(1, 2, \ldots, n)\Bigr|_{\rm cut}  =  \cr
& \hskip -3mm
\int\! {d^{4-2\eps}p\over (2\pi)^{4-2\eps}} \; {i\over {\ell_1^2-m_1^2} } \, 
A_{n_1}^\tree (-\ell_1, i_1, \ldots, i_2,\ell_2) \, 
{i\over {\ell_2^2-m_2^2}} \,A_{n_2}^\tree (-\ell_2 ,i_2+1,\ldots, i_1-1,\ell_1)
\biggr|_{\rm cut} \, .\cr}
\equn
\label{TreeProductDef}
$$
This equation is valid only for the cut channel under consideration.
The loop momentum is $p$; $\ell_1$ and $\ell_2$ are the momenta of the
cut propagators, with masses $m_1$ and $m_2$ respectively and
$n_1+n_2=n+4$.  One then reconstructs the complete amplitude by
finding a function which has the correct cuts in {\it all} channels.

%

Since the loop integral in \eqn{TreeProductDef} is composed of
products of on-shell tree amplitudes, the basic building blocks are
gauge invariant.  These tree amplitudes are inherently easier to
calculate than loop amplitudes and in many cases have been previously
calculated.  A key feature of the cutting method is that a specific
subset of scalar integral coefficients are obtained in any given
channel.  Then one repeats this process for various channels thus
computing the coefficients for all the scalar integrals.  There are
integrals that do not appear in any channel, namely the tadpole and
scalar bubble integrals,
$$
I^{D=4-2\eps}_1 \hskip 1 cm \mbox{and} \hskip 1 cm 
\lim_{k^2\rightarrow 0} I^{D=4-2\eps}_2(k^2) \, .
\equn
$$
However, with prior knowledge of the infrared and ultraviolet behavior
of an amplitude in many cases the coefficients of these integrals can
be fixed, as discussed in ref.~\cite{Massive}.

It is generally convenient to express the tree amplitudes in
\eqn{TreeProductDef} using the spinor helicity formalism discussed in
\sec{SpinorHelicity}. Then one can evaluate the right-hand side of
\eqn{TreeProductDef} by constructing spinor strings of the form
$\sand{a}.{\cdots \lsl_1 \cdots \lsl_2 \cdots}.{b}$, and then
commuting the $\lsl_i$ towards each other.  This generates terms of
the form $\ell_i \cdot k_j$, which may be rewritten as differences of
inverse propagators and terms which do not depend on the loop
momentum. This type of reduction procedure has been previously used in
a number of references \cite{SusyFour,SusyOne,Massive,Pittau}. We
shall make use of it in the next section.

%
\begin{figure}[ht]
\begin{center}
\epsfig{file=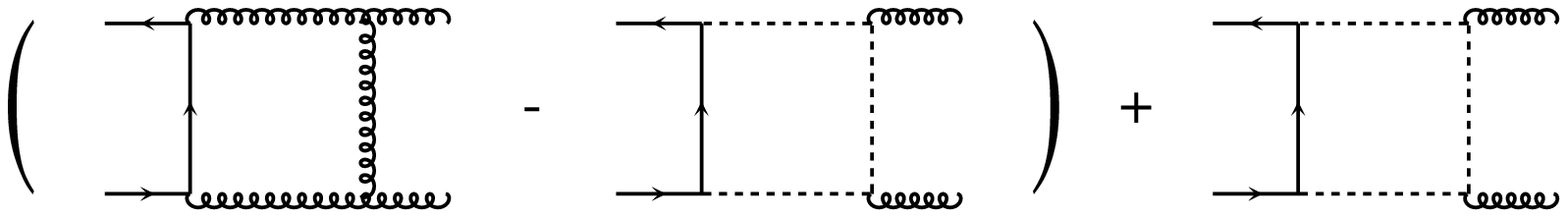,clip=,width=5.3in}
\end{center}
\vskip -.75 cm
\caption[]{The contribution from a vector in the loop is separated
into the difference of a vector and scalar plus a scalar so that the
difference satisfies the power-counting criterion.
\label{ScalarGluonFigure}
}
\end{figure}

When using spinor helicity one must take into account that the momenta
of the intermediate on-shell lines are in $(4-2\eps)$ dimensions while
the helicity formalism is defined in four dimensions.  If one drops
the $(-2\eps)$-dimensional components of the momenta, one could make
an error in rational functions arising from $\eps/\eps$ terms. For
five- and higher-point calculations it can be more convenient to use
four-dimensional momenta in the cuts, but then to reconstruct the
missing rational functions using collinear and multi-particle
factorization properties \cite{Factorization,Review}.  For four-point
calculations it is convenient to keep terms to all orders in~$\eps$,
since factorization is insufficient to fix all rational functions.

The cases of fermion or scalar lines with $(4-2\eps)$ momenta in the
helicity formalism have been extensively discussed in
ref.~\cite{Massive} and do not present any difficulties.  Although it
is unnecessary for the specific calculation $gg\rightarrow gH$, a
convenient strategy for handling channels with gluon lines is to
subtract and add equivalent diagrams with internal gluon lines
replaced by scalar or fermion lines, as depicted in
\fig{ScalarGluonFigure}. In some cases one uses fermions so as to
group particles together with supersymmetric partners. Such
combinations of diagrams satisfy the power counting criterion that
$n$-point tensor integrals should have (for $n>2$) at most $n-2$
powers of loop momentum in the numerator of the integrand; two-point
integrals should have at most one power of loop momentum.  The
required rearrangement of the diagrams to make the power count
apparent is discussed in ref.~\cite{SusyOne}.  In particular, for the
example portrayed in \fig{ScalarGluonFigure} one should use background
field Feynman gauge \cite{Background}.  The origin of this power
counting criterion is from the requirement that all contributions due
to the $\Ord(\eps)$ parts of loop momenta should not interfere with
ultraviolet poles to yield $\Ord(\eps^0)$ contributions.  (One can
also show that infrared divergences never interfere with the
$\Ord(\eps)$ parts of loop momenta \cite{SusyOne}.)  This criterion
also holds for the general massive amplitudes discussed in this paper.
This is useful since one can then take the sewed gluon lines to have
four-dimensional momenta without any errors through $\Ord(\eps^0)$.

\subsection{From Feynman diagrams to the unitarity method}

We now describe how to perform a Feynman diagram calculation so as to
mimic the cutting method.  In the next section we give an explicit
example involving both internal and external masses.  The basic idea
is rather simple.

Given that we know all the scalar integrals that could possibly enter
a one-loop amplitude, suppose we target those scalar integrals that
have a pair of scalar propagators
$$
i/(\ell_1^2-m_1^2)\hskip 1 cm   \mbox{and} \hskip 1 cm  i/(\ell_2^2-m_2^2)\, ,
\equn
$$ 
where $\ell_1$ is the momentum of the scalar propagator between
external legs $i_1-1$ and $i_1$, and $\ell_2$ is between $i_2$ and
$i_2+1$ of a general $n$-point amplitude, where the indicated
propagators are each of a specific particle type. In order to generate
a scalar integral with a specific propagator one must necessarily have
evaluated a Feynman diagram that has the corresponding propagator.

We can define a channel (this channel will correspond to the
cut-channel when the cut is performed on the indicated legs) as a
subset of the terms in the full amplitude; the terms being the scalar
integrals which include the pair of scalar propagators mentioned above
together with their corresponding coefficients, the $c_i$'s.

In each Feynman integral (the tensor integral corresponding to each
Feynman diagram) we explicitly extract a denominator that includes the
pair of scalar propagators under consideration. Then if we impose the
on-shell conditions $\ell_1^2-m_1^2=\ell_2^2-m_2^2=0$ to the
numerators of all the Feynman integrals contributing to the amplitude
then the result would yield exactly what we defined as a channel in
the previous paragraph.  The potential errors when imposing these
on-shell conditions would necessarily be of the form
$$
\int {d^{D} \ell \over (2\pi)^{D}} \, {{
(\ell_1^2-m_1^2)^n (\ell_2^2-m_2^2)^m  f(\ell_1,\ell_2,\ldots) } \over 
{ \cdots (\ell_1^2-m_1^2) (\ell_2^2-m_2^2) \cdots } } \, ,
\equn
\label{NotInChannel}
$$
where $n\geq 1$ and/or $m \geq 1$, and $D=4-2\eps$. Since
$(\ell_1^2-m_1^2)$ and $(\ell_2^2-m_2^2)$ are themselves inverse
scalar propagators, anything of the form of \eqn{NotInChannel} could
only generate terms which are not in the channel under
consideration. This is because one or both of the inverse propagators
would cancel a scalar propagator in the denominator and generate a
scalar integral not included in the specified channel, i.e. that does
not include both of the indicated scalar propagators.  Setting
$\ell_1^2-m_1^2=\ell_2^2-m_2^2=0$ only affects the calculation of
those terms not in the specified channel.  Thus the evaluation of the
$c_i$'s that are in the specified channel are not affected by imposing
the on-shell condition.

By systematically stepping through a sufficient number of channels one
can fix all the coefficients, the $c_i$'s in \eqn{temp1}.  In this way
we can reconstruct the full amplitude.  For some coefficients the
corresponding integrals appear in multiple channels; for consistency
these must coincide.

\begin{figure}[ht]
\begin{center}
\epsfig{file=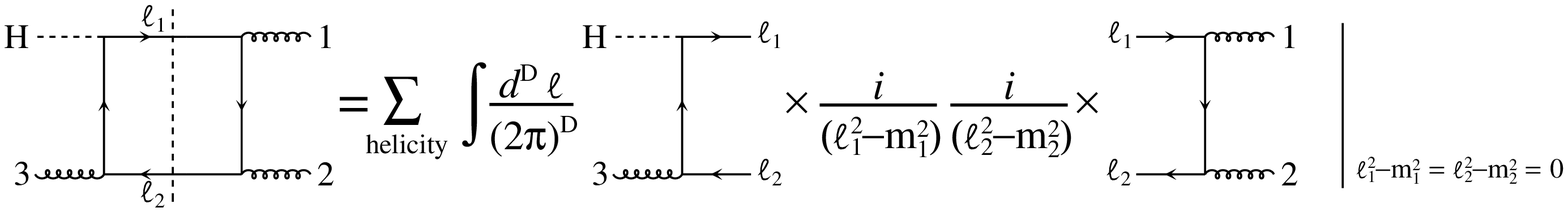,clip=,width=6.5in}
\end{center}
\vskip -.75 cm
\caption[]{
\label{Factorize}
An example of a Feynman graph for the $gg\rightarrow gH$ amplitude that 
contributes to the $s$-channel, `factorized' into the sewing of two trees. 
The dashed lines on the left hand side indicate the on-shell propagators
and the helicity sum is over the polarizations of the `factorized' legs.
}
\end{figure}

So how does using the on-shell conditions $\ell_1^2-m_1^2=\ell_2^2-m_2^2=0$ 
in the numerator help us in evaluating the Feynman diagrams?
All Feynman diagrams in a specific channel can now be `factorized'
using their on-shell legs into the sewing of two Feynman tree diagrams 
(see \fig{Factorize}).
%

In figure \fig{Factorize} the `factorized' Feynman graph only contains
intermediate on-shell fermions for the $gg\rightarrow gH$
amplitude. But in general this `factorization' process can be
performed on any one-loop Feynman diagram. With on-shell scalars or
fermions it is clear that this can be done. With intermediate massless
gauge particles one might worry that one has forgotten about the
contributions of diagrams with ghosts on the right hand side of
\fig{Factorize}. We know, however, that the net effect of the one-loop
diagrams with ghosts is to cancel the unphysical polarizations of
graphs with corresponding gauge particles.  This cancelation can be
made apparent using the background field Feynman gauge
\cite{Background}. However, once one has `factorized' a Feynman graph
on a pair of propagators that includes a gauge particle then the
contribution of the ghosts is not necessary as the sewing of the two
on-shell Feynman trees manifestly has no unphysical gauge degrees of
freedom.

If one sums over all the Feynman diagrams that contribute to a
specific channel and we impose the on-shell condition then the
equivalent equation to \fig{Factorize} would have on-shell tree
amplitudes sewed together instead of on-shell Feynman trees:
$$
\eqalign{
\left. A_n(1,2,\ldots,n)
\right|_{\rm channel}
= & \sum_{\rm helicity} \int {d^{4-2\epsilon} \ell \over (2\pi)^{4-2\epsilon}}
A^{\rm tree}_{n_1}(-\ell_1,i_1,\ldots,i_2,\ell_2) 
{i \over (\ell_1^2-m_1^2)} {i \over (\ell_2^2-m_2^2)}\cr
& \hskip 3.5cm
\times 
A^{\rm tree}_{n_2}(-\ell_2,i_2+1,\ldots,i_1-1,\ell_1) 
\Bigr|_{\ell_1^2-m_1^2=\ell_2^2-m_2^2=0} \, .
}
\equn
\label{SewedAmplitudes}
$$

The sum on the right-hand side runs over all helicity states that can
propagate across the `cut' legs. Using the above equation one can
compute all the coefficients of the scalar integrals that appear in
this channel.

The above equation, \eqn{SewedAmplitudes}, is the unitarity method
mentioned earlier in this section. Originally the method was inspired
by Cutkosky rules, however, from the arguments presented here it is
clear that one can evaluate Feynman diagrams in a sophisticated manner
so as to take advantage of the benefits inherent in the unitarity
method.  Instead of performing the calculation of the scalar integral
coefficients in parallel, we perform the calculation by targeting
simultaneously only those integrals in a given channel.

The notion of channels presented in this section is more general than
that of just considering Cutkosky rules because we may use the
on-shell conditions on any propagator which does not collapse in the
integral function under consideration.  (Indeed, this observation has
already been used in calculations of $Z \rightarrow 4$ partons
\cite{Zjets}.) In general we could construct a channel where we use
the on-shell condition for any number of propagators (it is even
possible to construct a channel where the on-shell condition is used
for a single propagator); however, it is normally convenient to use
only the two-propagator channel construction.

\section{Calculation of $gg  \rightarrow gH$}
\label{Calculation}

\begin{figure}[ht]
\begin{center}
\epsfig{file=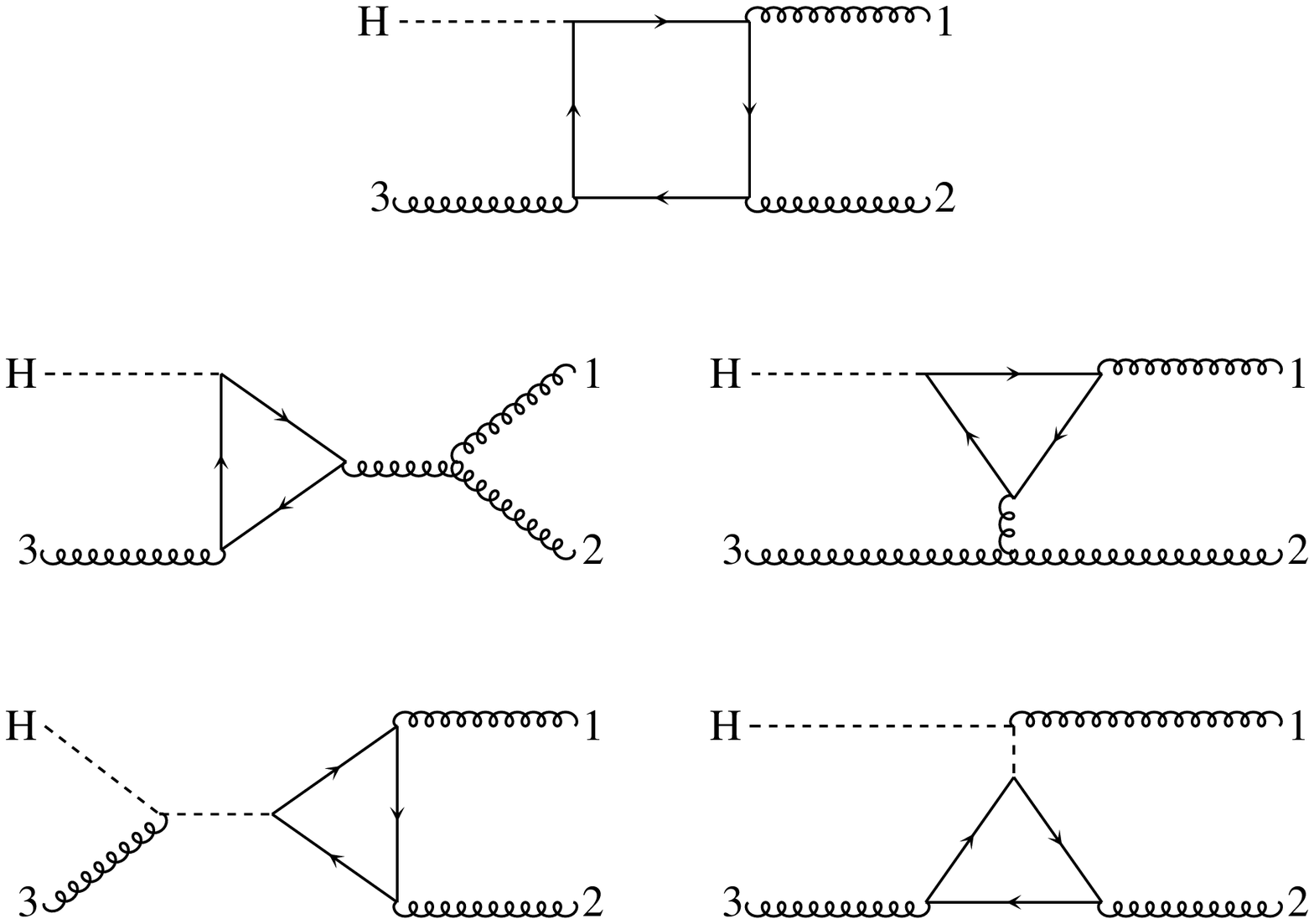,clip=,width=4.5in,height=3in}
\end{center}
\vskip -.75 cm
\caption[]{
\label{FeynmanDiagrams}
The color-ordered Feynman diagrams contributing to the amplitude
$gg\rightarrow gH$.  }
\end{figure}

In this section we apply the procedure outlined in the previous
section to the specific example of $gg \rightarrow gH$.  This
amplitude has one massive external leg and all the Feynman diagrams
have a massive quark running in the loop. See \fig{FeynmanDiagrams}
for the color-ordered Feynman diagrams contributing to this amplitude.
%

In \fig{FeynmanDiagrams} there are two Feynman diagrams that involve a
Higgs-Higgs-gluon coupling which does not exist in the Standard
Model. As mentioned in \sec{ColorDecomposition}, although these
diagrams cancel from the final result, they make the primitive
amplitudes (which have a fixed ordering of external legs) gauge
invariant.  Of course, one does not need to include the extra
diagrams, but then one would need to sum over the permutations with
all possible external orderings of the Higgs, as in
\eqn{partialdecomposition}, to recover gauge invariance.

\begin{figure}[ht]
\begin{center}
\epsfig{file=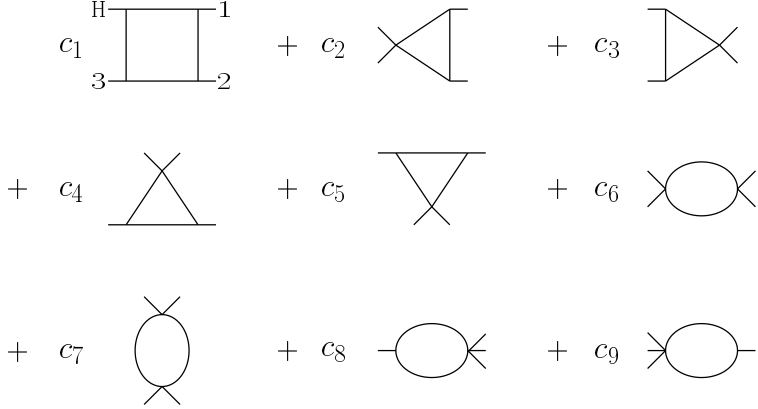,clip=,width=4in}
\end{center}
\vskip -.75 cm
\caption[]{
\label{ScalarIntegrals}
The $gg\rightarrow gH$ amplitude parameterized in terms of all the
scalar integrals that might enter. The external legs are labeled
clockwise starting at the top right.  }
\end{figure}

As mentioned in \sec{Feynman} by observing the Feynman diagrams in
\fig{FeynmanDiagrams} we can write down all the scalar integrals that
can enter the amplitude, see \fig{ScalarIntegrals}.
%
The integrals are two-, three- and four-point scalar integrals that
have massive internal propagators and some combination of massive
external kinematics. One-point tadpole integrals do not appear since
they are forbidden by power counting. (When using Passarino-Veltman
reduction one could encounter such integrals, but they must cancel in
the final expressions for this amplitude since one does not encounter
them when using Feynman parameterization.) The scalar integrals are
found in \app{ScalarIntegralFunctions}.

In addition to the use of color decomposition,
\sec{ColorDecomposition}, it is also convenient to decompose the
primitive amplitudes into all possible external helicity
configurations. So the objects that need to be computed are the color
ordered gauge invariant primitive amplitudes. There are three such
objects; they are $A_4(1^+,2^+,3^+,H)$, $A_4(1^+,2^+,3^-,H)$ and
$A_4(1^+,2^-,3^+,H)$. The other primitive amplitudes (e.g.
$A_4(1^-,2^-,3^+,H)$) can be obtained from the above three primitive
amplitudes using parity and relabelings.

Since this amplitude has already been calculated by Ellis
et~al. \cite{Ellis} the purpose of the calculation is to explicitly
demonstrate that we may obtain the correct results by applying the
unitarity method to massive amplitudes, starting from Feynman
diagrams.  For illustrative purposes, only the calculation of
$A_4(1^+,2^+,3^+,H)$ is explicitly presented here -- the results of
the calculation of the other two primitive amplitudes are shown for
completeness.

\subsection{The $A_4(1^+,2^+,3^+,H)$ primitive amplitude}
\label{ppp}

\begin{figure}[ht]
  \begin{center}
  \vskip -0.5 cm
    ~\epsfig{file=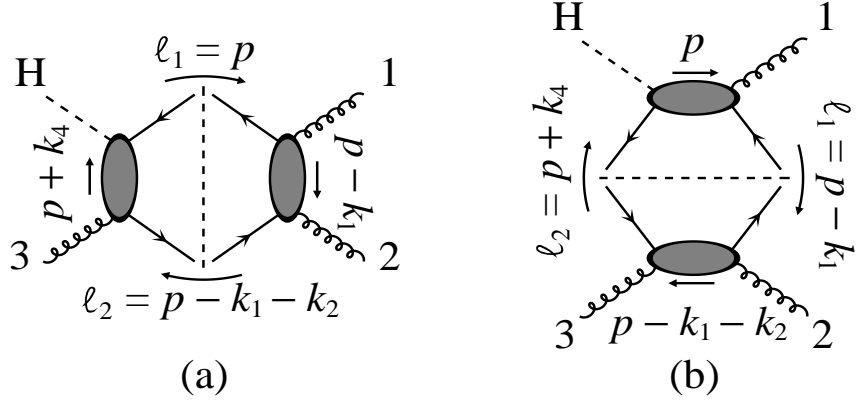,width=12cm,height=6cm,clip=}
    \caption[]{
      \label{sandtcuts}
  The $s$- and $t$-channels of the $gg\rightarrow gH$ 
amplitude. The on-shell lines are fermions since at tree level quarks 
are the only particles to couple both to gluons and to the scalar 
Higgs boson.
}
  \end{center}
\end{figure}

In order to compute the primitive amplitude $A_4(1^+,2^+,3^+,H)$ we
will construct it from multiple channels. We will see that it is
sufficient to only compute the $s$- and $t$-channels. The Mandelstam
variables are defined as $s\equiv (k_1+k_2)^2$ and $t\equiv
(k_2+k_3)^2$. These channels can be seen in \fig{sandtcuts}.

We begin by first constructing the $s$-channel by sewing together the
two tree amplitudes on the left- and right-hand side of
\fig{sandtcuts}a. This corresponds to targeting the coefficients,
$c_1$, $c_2$, $c_3$ and $c_6$ in \fig{ScalarIntegrals}.  The tree
amplitude on the right of \fig{sandtcuts}a is given by \cite{Massive}
$$
A_4^{\rm tree}(-L_1,1^+,2^+,L_2) = i \frac{\bkpm{12}}{\bkmp{12}} 
\frac{\langle -L_1 | \omega_+(\s{\mu}+m) | -L_2 \rangle }
{(\ell_1-k_1)^2-\mu^2-m^2} \, ,
\equn\label{firsttree}
$$ 
where $\omega_\pm=\frac{1}{2}(1\pm\gamma_5)$ is the helicity
projection operator and the tree amplitude on the left is given by
$$
A_4^{\rm tree}(-L_2,3^+,H,L_1) = -\frac{im}{\sqrt{2}v} \left(
\frac{\langle -L_2 | \, \s{\pol}_3^+(\s{k}_4-2m) | -L_1 \rangle }
{(\ell_1+k_4)^2-\mu^2-m^2} + \frac{2k_4 
\cdot\pol_3^+}{s-m_H^2} \langle -L_2 | -L_1 \rangle  \right) \, . 
\equn\label{Higgstree}
$$
Here $v=\frac{2m_W}{g_W}$ is the vacuum expectation value of the
scalar Higgs field \cite{PeskinSchroeder}, $m$ is the mass of the
fermion in the loop and $m_H$ is the scalar Higgs mass.  Capitalized
momenta are $(4-2\eps)$-dimensional, lower case momenta are
four-dimensional and $\mu$ is the $(-2\epsilon)$-dimensional part of
the loop momentum. (See \app{IntegralsAppendix} and \app{MuSlashEtc}.)
In these tree amplitudes we follow the conventions of
ref.~\cite{Massive} for massive spinors, summarized in
\app{MuSlashEtc}. In this convention the distinction between fermion
and anti-fermion spinors is suppressed.

As depicted in \fig{sandtcuts}a, for the $s$-channel the momenta of
the on-shell legs are
$$
\ell_1=p \, , \hskip 2.5 cm \ell_2=p-k_1-k_2 \, .
\equn
$$
The second tree amplitude, \eqn{Higgstree}, involving the Higgs has 
been constructed from the tree level Feynman diagrams including the
additional fictitious Higgs-Higgs-gluon coupling (this is the standard 
scalar-scalar-vector coupling of scalar QED \cite{PeskinSchroeder})
mentioned earlier.

So sewing the two tree amplitudes together gives us the $s$-channel of
the amplitude
$$
\eqalign{
A_4(1^+,2^+,3^+,H)\Bigr|_{s-{\rm channel}} = & 
- \frac{\sqrt{2} m \, k_4\cdot\pol_3^+}{v(s-m_H^2)} 
\frac{\bkpm{12}}{\bkmp{12}} 
\int \frac{d^{4-2\epsilon}P}{(2\pi)^{4-2\epsilon}} \frac{\N_1}{\D_3}
\cr
& \left. \hskip 2.5cm -\frac{m}{\sqrt{2}v} \frac{\bkpm{12}}{\bkmp{12}} 
\int \frac{d^{4-2\epsilon}P}{(2\pi)^{4-2\epsilon}}
\frac{\N_2}{\D_4} 
\right|_{
\begin{array}{c}
{\scriptstyle \ell_1^2-\mu^2-m^2=0} \cr
{\scriptstyle \ell_2^2-\mu^2-m^2=0}
\end{array}
}
\, , }
\equn\label{ScutIntegral}
$$
where the denominators are 
$$
\eqalign{
\D_3 = \: & [\ell_1^2-\mu^2-m^2][(\ell_1-k_1)^2-\mu^2-m^2]
[(\ell_1-k_1-k_2)^2-\mu^2-m^2] \cr
\D_4 = \: & [\ell_1^2-\mu^2-m^2][(\ell_1-k_1)^2-\mu^2-m^2]
[(\ell_1-k_1-k_2)^2-\mu^2-m^2][(\ell_1+k_4)^2-\mu^2-m^2] \, , }
\equn
$$ 
and the numerators are
$$
\eqalign{
\N_1 = \: & \langle   -L_1 | \omega_+(\s{\mu}+m) | -L_2 \rangle \langle -L_2 |
 -L_1 \rangle \cr
\N_2 = \: & \langle   -L_1 | \omega_+(\s{\mu}+m) | -L_2 \rangle \langle -L_2 |
\s{\pol}_3^+(\s{k}_4-2m)  | -L_1 \rangle \, . }
\equn
$$
The two terms in \eqn{ScutIntegral} correspond to the two terms in 
\eqn{Higgstree}.

First, concentrating on the numerator $\N_1$ we can write it in the 
form of a trace
$$
\N_1 = \tr_+ [(\s{\mu}+m)(\s{\ell}_2+\s{\mu}-m)(\s{\ell}_1+\s{\mu}-m)] \, ,
\equn
$$
where we use the notation $\tr_\pm[\ldots]=\tr[\omega_\pm\ldots]$.
When 
expanding out the trace we first concentrate on the $\s{\mu}$'s since 
the final expression for the numerator can only depend on $\mu^2$ 
since the $(-2\epsilon)$-dimensional integral over $\mu$ vanishes for 
a integrand with a odd number of $\sl{\mu}$'s. Thus,
$$
\eqalign{
\N_1 = \; & \tr_+ [(m\s{\ell}_2-\mu^2-m^2)(\s{\ell}_1-m)] \cr
     = \; & 2m(m^2+\mu^2) + 2m \, \ell_1\cdot\ell_2 \cr
     = \; & 4m(m^2+\mu^2) - m s \, . \cr }
\equn
$$

The numerator $\N_2$ can also be written in the form of a trace 
$$
\eqalign{
\N_2 = \;& \tr_+ [(\s{\mu}+m)(\s{\ell}_2+\s{\mu}-m)\s{\pol}_3^+(\s{k}_4-2m)
(\s{\ell}_1+\s{\mu}-m)] \, . \cr }
\equn
$$
So again first concentrating on the $\s{\mu}$'s we can write
$$
\eqalign{
\hskip .5 cm
\N_2 = \; & \tr_+ [(\s{\mu}\s{\ell}_2+m\s{\ell}_2-\mu^2-m^2)\s{\pol}_3^+
(\s{k}_4-2m)(\s{\ell}_1+\s{\mu}-m)] \cr
    = \; & \tr_+ [(m\s{\ell}_2-\mu^2-m^2)\s{\pol}_3^+(\s{k}_4-2m)
(\s{\ell}_1-m)]-\mu^2 \, \tr_+ [\s{\ell}_2\s{\pol}_3^+(\s{k}_4-2m)]
 \cr
    = \; & 2m(m^2+\mu^2)(2\ell_1+2\ell_2+k_4)\cdot\pol_3^+ + 
m \, \tr_+[\s{\ell}_2\s{\pol}_3^+\s{k}_4\s{\ell}_1] \, . }
\equn
$$
Recalling that $\ell_2=\ell_1-k_1-k_2$ and choosing the reference momenta 
$q_3=k_1$ (so that $k_1.\pol_3^+=0$) then 
$$
\N_2 = 8m(m^2+\mu^2)\ell_1\cdot\pol_3^+ + 6m(m^2+\mu^2)k_4\cdot
\pol_3^++ m \, \tr_+ [\s{\ell}_2\s{\pol}_3^+\s{k}_4\s{\ell}_1] \, .
\equn
$$
We can evaluate the remaining trace in the expression for $\N_2$ by using 
\eqn{PolSlash}
$$
\tr_+ [\s{\ell}_2\s{\pol}_3^+\s{k}_4\s{\ell}_1] = 
\tr_- [\s{\pol}_3^+\s{k}_4\s{\ell}_1\s{\ell}_2] =
\frac{\sqrt{2}}{\bkmp{13}}\langle 1^- | 4 \ell_1 \ell_2 | 3^-\rangle \, ,
\equn
$$
where for brevity we represent external momenta by their indices and
neglect slashing the momenta inside the inner-product. Using
$\s{\ell}_1\s{\ell}_2=\s{\ell}_1(\s{\ell}_1-\s{k}_1-\s{k}_2)=m^2+\mu^2-
\s{\ell}_1(\s{k}_1+\s{k}_2)$ and momentum conservation the trace can
be expressed as
$$
\tr_+ [\s{\ell}_2\s{\pol}_3^+\s{k}_4\s{\ell}_1] = 
\frac{\sqrt{2}}{\bkmp{13}}\left\{ (m^2+\mu^2)\langle1^-|4|3^-\rangle + 
\langle 1^-|4\ell_14|3^-\rangle \right\} \, .
\equn
$$
But since
$$
k_4\cdot\pol_3^+ = \frac{\langle 1^-|4|3^- \rangle}
{\sqrt{2}\bkmp{13}} \, , 
\equn\label{innerproduct}
$$
$\N_2$ can be expressed as
$$
\N_2= 8m(m^2+\mu^2)(\ell_1\cdot\pol_3^++k_4\cdot\pol_3^+) + 
\sqrt{2}m\frac{\langle 1^-|4\ell_14|3^-\rangle}{\bkmp{13}} \, .
\equn
$$

The integral comprised of the numerator $\N_2$ now only involves at
most a single power of loop momentum $\ell$ in the integrand. This can
be readily evaluated with the use of Feynman parameterization.  The
Feynman parameter shift is
$$
\ell_1 = q + a_2 k_1 + a_3 (k_1+k_2) -a_4 k_4 \, ,
\equn
$$ 
where $q$ is the shifted loop momentum. After Feynman parameterizing
the integral the terms in $\N_2$ involving $\ell_1$ can be evaluated
by performing the momentum shift and next dropping terms containing a
single power of $q$ which vanish. So the two terms in $\N_2$ involving
$\ell_1$ become
$$
\ell_1.\pol_3^+ = \frac{ \langle1^-| q + a_2 k_1 + a_3 (k_1+k_2) -a_4 k_4
|3^-\rangle }{\sqrt{2}\bkmp{13}} = -(a_3+a_4) k_4.\pol_3^+ \, ,
\equn
$$
and
$$
\eqalign{
\hskip 1cm
\langle 1^-|4\ell_14|3^-\rangle = \; & a_2 \langle 1^-|414|3^-\rangle
+a_3 \langle 1^-|4(1+2)4|3^-\rangle - a_4 \langle 1^-|444|3^-\rangle \cr
= \; & [(t-m_H^2)a_2 -sa_3 -m_H^2a_4]\langle 1^-|4|3^-\rangle \, , }
\equn
$$
after performing the momentum shift. 

As defined in \app{FeynmanParamAppendix}, the integral function 
$I_n[a_i]$ is a Feynman parameterized form of the loop integral 
with a factor of $a_i$ in the numerator. Thus after Feynman 
parameterization \eqn{ScutIntegral} becomes 
$$
\eqalign{
A_4(1^+,2^+,3^+,H)\Bigr|_{s-{\rm channel}} = & -i \frac{\sqrt{2}m^2
k_4\cdot\pol_3^+}{v(4\pi)^{2-\epsilon}}\frac{\bkpm{12}}{\bkmp{12}}
\Biggl\{ \frac{1}{s-m_H^2} I_3^{(4)}[4(m^2+\mu^2)-s]
- (t-m_H^2)I_4[a_2] \cr
& +sI_4[a_3] + m_H^2I_4[a_4] -4I_4[(m^2+\mu^2)(1-a_3-a_4)] 
\Biggr\}\Biggr|_{s-{\rm channel}} \, , }
\equn
$$
where $I^{D=4-2\epsilon}_n[\mu^2a_i]=-\epsilon
I^{D=6-2\epsilon}_n[a_i]$ and $I_3^{(j)}$ is defined in
\app{IntegralsAppendix}.  The dimension label on
$(4-2\epsilon)$-dimensional integrals has been suppressed. Then using
the recursive formula for $I_n[a_i]$ in \eqn{LinearIntegralRelation}
and \eqn{innerproduct} one can write the $s$-channel of the primitive
amplitude as
$$
\eqalign{
A_4(1^+,2^+,3^+,H)\Bigr|_{s-{\rm channel}} = & - \frac{i}{(4\pi)^{2-\epsilon}}
\frac{m^2}{v} \frac{st}{\bkmp{12}\bkmp{23}\bkmp{31}} 
\Biggl\{ \frac{t-u}{2t(s-m_H^2)}I_3^{(4)}[4(m^2+\mu^2)-m_H^2] \cr
& \hskip -1.3cm -\frac{s-m_H^2}{2st}I_3^{(2)}[4(m^2+\mu^2)-m_H^2] 
-\frac{1}{2}I_4[4(m^2+\mu^2)-m_H^2] \Biggr\}\Biggr|_{s-{\rm channel}} \, , 
}
\equn
$$
where $u=m_H^2-s-t$, also noting that $I_3^{(2)}$ is a scalar triangle
integral dependent on the kinematic invariant $s$ and $m_H^2$, while
$I_3^{(4)}$ is a function of $s$ only.  Then using \eqn{DSEIntegrand}
the $s$-channel becomes
$$
\eqalign{
& A_4(1^+,2^+,3^+,H)\Bigr|_{s-{\rm channel}} = 
- \frac{i}{(4\pi)^{2-\epsilon}}
\frac{m^2}{v} \frac{st}{\bkmp{12}\bkmp{23}\bkmp{31}} 
\Biggl\{(4m^2-m_H^2)\Biggl[\frac{t-u}{2t(s-m_H^2)}I_3^{(4)} 
\cr
& \hskip 0 cm
-\frac{s-m_H^2}{2st}I_3^{(2)} -\frac{1}{2}I_4\Biggr] -4\epsilon\Biggl[
\frac{t-u}{2t(s-m_H^2)}I_3^{(4)D=6-2\epsilon}
-\frac{s-m_H^2}{2st}I_3^{(2)D=6-2\epsilon}-\frac{1}{2}
I_4^{D=6-2\epsilon} \Biggr] \Biggr\} \Biggr|_{s-{\rm channel}} \, . }
\equn\label{ScutAnswer}
$$

Now that we have the $s$-channel of this amplitude we need to
calculate the $t$-channel (corresponding to targeting $c_1$, $c_4$,
$c_5$ and $c_7$ in \fig{ScalarIntegrals}) in order to reconstruct the
full primitive amplitude.  However due to the symmetry of the all-plus
primitive amplitude it is not necessary to explicitly perform this
calculation as the $s$- and $t$-channels are related up to an overall
sign by a simple exchange of legs $1$ and $3$ which is equivalent to
exchanging $s$ and $t$ in \eqn{ScutAnswer} (and also switching the $1$
and $3$ in the inner products in the pre-factor of the
expression). Thus the $t$-channel of $A_4(1^+,2^+,3^+,H)$ is given by
$$
\eqalign{
& A_4(1^+,2^+,3^+,H)\Bigr|_{t-{\rm channel}} = - \frac{i}{(4\pi)^{2-\epsilon}}
\frac{m^2}{v} \frac{st}{\bkmp{12}\bkmp{23}\bkmp{31}} 
\Biggl\{(4m^2-m_H^2)\Biggl[\frac{s-u}{2s(t-m_H^2)}I_3^{(1)}
\cr
& \hskip 0 cm -\frac{t-m_H^2}{2st}I_3^{(3)}
-\frac{1}{2}I_4\Biggr]-4\epsilon\Biggl[
\frac{s-u}{2s(t-m_H^2)}I_3^{(1)D=6-2\epsilon}
-\frac{t-m_H^2}{2st}I_3^{(3)D=6-2\epsilon}-\frac{1}{2}
I_4^{D=6-2\epsilon} \Biggr] \Biggr\} \Biggr|_{t-{\rm channel}} \, , }
\equn\label{TcutAnswer}
$$
again recalling that $I_3^{(3)}$ is a scalar triangle integral
dependent on the kinematic invariant $t$ and $m_H^2$, while
$I_3^{(1)}$ is a function of $t$ only. It is also useful to observe
that the coefficient of the $I_4$ integral in the $t$-channel matches
that of the $s$-channel as expected.  If we refer back to
\fig{ScalarIntegrals}, where all the scalar integrals that could
possibly enter were listed, we notice that the coefficients, $c_8$ of
$I_2(m_H^2)$ and $c_9$ of $I_2(0)$ have not yet been fixed.  However,
there are no other UV infinities in either the $s$- or $t$-channel
expressions (eqs.~(\ref{ScutAnswer}) and (\ref{TcutAnswer})), so
therefore since this primitive amplitude is UV-finite (since it
vanishes at tree level), it is not surprising that these coefficients
are zero.  This may be explicitly verified by direct construction of
the channel corresponding to using the on-shell condition on the two
propagators adjacent to the massive external leg, showing that both
$c_8$ and $c_9$ are individually zero. We have not evaluated this
channel, since this example is only for illustrative purposes.

Thus with the two channels of the primitive amplitude we can 
construct the full primitive amplitude
$A_4(1^+,2^+,3^+,H)$, which is
$$
\eqalign{
& A_4(1^+,2^+,3^+,H) = - \frac{i}{(4\pi)^{2-\epsilon}}
\frac{m^2}{v} \frac{st}{\bkmp{12}\bkmp{23}\bkmp{31}} 
\Biggl\{(4m^2-m_H^2)\Biggl[\frac{s-u}{2s(t-m_H^2)}I_3^{(1)}
-\frac{s-m_H^2}{2st}I_3^{(2)} \cr
& \hskip 2 cm -\frac{t-m_H^2}{2st}I_3^{(3)}
+ \frac{t-u}{2t(s-m_H^2)}I_3^{(4)}-\frac{1}{2}I_4\Biggr] 
-4\epsilon\Biggl[\frac{s-u}{2s(t-m_H^2)}I_3^{(1)D=6-2\epsilon} \cr
& \hskip 2 cm -\frac{s-m_H^2}{2st}I_3^{(2)D=6-2\epsilon}
-\frac{t-m_H^2}{2st}I_3^{(3)D=6-2\epsilon}
+\frac{t-u}{2t(s-m_H^2)}I_3^{(4)D=6-2\epsilon}-\frac{1}{2}
I_4^{D=6-2\epsilon} \Biggr] \Biggr\} \, . }
\equn\label{PPPAmplitude}
$$
Note that this expression is valid to all orders in $\eps$ with
four-dimensional external momenta.  Since this amplitude has no
ultraviolet or quadratic divergences, the primitive amplitude is well
defined in the limit $\epsilon\rightarrow 0$. Therefore
\eqn{PPPAmplitude} can be simplified, since in this limit with the use
of \eqn{HigherToLower}, $\epsilon
I_3^{D=6-2\epsilon}\rightarrow\frac{1}{2}$ and $\epsilon
I_4^{D=6-2\epsilon}\rightarrow 0$. Also with the intention of
simplifying this expression we can use \eqn{integral2invariants}
explicitly. Thus
$$
\eqalign{
A_4(1^+,2^+,3^+,H) = & - \frac{i}{(4\pi)^2}
\frac{m^2}{v} \frac{st}{\bkmp{12}\bkmp{23}\bkmp{31}} 
\Biggl\{(4m^2-m_H^2)\Biggl[\frac{1}{s-m_H^2}I_3(s)
+\frac{1}{t-m_H^2}I_3(t)  \cr
& +\frac{m_H^2}{st}I_3(m_H^2) -\frac{1}{2}I_4(s,t,m_H^2)\Biggr] 
-\frac{2u(m_H^4-ts)}{ts(s-m_H^2)(t-m_H^2)} \Biggr\} \, , }
\equn\label{answer1}
$$
where all the integrals in this equation are in four dimensions.

\subsection{The remaining primitive amplitudes}

In a similar fashion to the all-plus helicity case the primitive
amplitudes corresponding to the remaining two helicity configurations
may be calculated.  The results are
$$
\eqalign{
& A_4(1^+,2^+,3^-,H) = - \frac{i}{(4\pi)^2} \frac{m^2}{v}
\frac{\bkpm{12}^3}{\bkpm{13}\bkpm{23}} \frac{t}{s} \Biggl\{
\frac{4um^2}{(t-m_H^2)^2}I_3(t) 
- \frac{4um^2m_H^2}{t(t-m_H^2)^2}I_3(m_H^2) \cr
& \hskip 2 cm + (4m^2-s)\Biggl[ \frac{1}{s-m_H^2}I_3(s) 
+ \frac{1}{t-m_H^2}I_3(t) -\frac{m_H^2}{t(t-m_H^2)}I_3(m_H^2) 
- \frac{1}{2}I_4(s,t,m_H^2) \Biggr] \cr
& \hskip 2 cm - \frac{2u}{(t-m_H^2)^2}\left[\tilde{I}_2(t) 
-\tilde{I}_2(m_H^2)\right]
+ \frac{2u(t-s)}{t(s-m_H^2)(t-m_H^2)}\Biggr\} \, , }
\equn\label{answer2}
$$
and
$$
\eqalign{
& A_4(1^+,2^-,3^+,H) = - \frac{i}{(4\pi)^2} \frac{m^2}{v}
\frac{\bkpm{13}^3}{\bkpm{12}\bkpm{23}} \frac{st}{u^2} \Biggl\{
-\frac{1}{2}\Biggl[u-12m^2-\frac{4ts}{u}\Biggr]I_4(s,t,m_H^2) \cr
& \hskip 1.0 cm
+\Biggl[\frac{u-4m^2}{s-m_H^2}+\frac{4s}{u}+\frac{4um^2}{(s-m_H^2)^2}
\Biggr]I_3(s)
+\Biggl[\frac{u-4m^2}{t-m_H^2}+\frac{4t}{u}+\frac{4um^2}{(t-m_H^2)^2}
\Biggr]I_3(t) \cr
& \hskip 1.0 cm
+m_H^2\Biggl[\frac{(u-4m^2)(st-u^2)}{st(s-m_H^2)(t-m_H^2)}-\frac{4}{u}
-\frac{4um^2}{s(s-m_H^2)^2}-\frac{4um^2}{t(t-m_H^2)^2}\Biggr]I_3(m_H^2) \cr
& \hskip 1.0 cm 
-\frac{2(3u+2t)}{(s-m_H^2)^2}\left[\tilde{I}_2(s)
-\tilde{I}_2(m_H^2)\right]
-\frac{2(3u+2s)}{(t-m_H^2)^2}\left[\tilde{I}_2(t)
-\tilde{I}_2(m_H^2)\right] -\frac{2u(u^2-st)}{st(s-m_H^2)(t-m_H^2)} 
\Biggr\} \, , }
\equn\label{answer3}
$$
where the $\tilde{I}_2$'s are bubble integrals with their $1/\epsilon$
poles removed (since the poles cancel in the difference between the
two $I_2$'s) and all integrals are in four dimensions.

Thus in equations (\ref{answer1},\ref{answer2},\ref{answer3}) we have
the three necessary primitive amplitudes $A_4(1^+,2^+,3^+,H)$,
$A_4(1^+,2^+,3^-,H)$ and $A_4(1^+,2^-,3^+,H)$ which are needed for the
full $gg\rightarrow gH$ amplitude. We have compared these amplitudes
to those of Ellis~et~al. \cite{Ellis} and they agree.  This comparison
is performed by dotting the tensor expressions of Ellis~et~al. with
spinor helicity polarization vectors.

\section{Summary and discussion}

In this paper we have explained the relationship of the unitarity
method reviewed in ref.\cite{Review} to Feynman diagram calculations.
One may calculate Feynman diagrams in a sophisticated manner so as to
take advantage of the benefits of the unitarity method.  We have
illustrated the usefulness of this method for the case of
$gg\rightarrow gH$.  This example is general in that it incorporates
both massive internal and external kinematics as well as a combination
of colored and colorless external particles. It should be evident from
the explicit calculation presented in this paper that this method
lends itself to making complicated analytical calculations more
tractable. This is because in this approach one is always dealing with
gauge invariant constructions, which are generally significantly
simpler than the gauge dependent Feynman graphs in a conventional
calculation.  In the unitarity method one targets the coefficients of
a subset of scalar integrals contributing to the amplitude by
computing a gauge invariant combination of Feynman diagrams.

The notion of a channel as defined in this paper is more general than
just the cut legs of a Cutkosky calculation, where one is nominally
computing the discontinuity of an amplitude in a certain cut
channel. The reason it is more general is that one defines a channel
as a subset of the terms in the full amplitude; those terms being the
scalar integrals that include some pre-defined combination of internal
propagators. This notion of a channel has already been useful in the
evaluation of $Z\rightarrow 4 \ \mbox{partons}$ \cite{Zjets}.  It was
also helpful in two-loop calculations in $N=4$ supersymmetric
Yang-Mills \cite{BernYanRoz}.

One of the themes of this paper has been to explain the usefulness of
the unitarity method to multi-point one-loop calculations. The ideas
discussed here will potentially be beneficial to the extension of
these types of calculations to multi-loops. Initial steps have already
been made \cite{BernYanRoz}, and more attention is being devoted to
these pursuits.

\vskip .5 cm 

\noindent
{\large \bf Acknowledgments}

I thank Z. Bern for encouragement and many valuable discussions.  I
also thank L. Dixon, A.K. Grant, D.A. Kosower, E. Marcus, A.G. Morgan
and B. Yan for helpful discussions and comments. This work was
supported by the DOE under contract DE-FG03-91ER40662.

\appendix

\section{Integrals}
\label{IntegralsAppendix}
In this appendix we collect expressions for the integrals
\cite{PV,Integrals,IntegralRecursion} we use in this paper.  The
$D$-dimensional loop integrals considered in this paper are defined by
$$
  I_n^{D}[P^{\alpha_1} \cdots P^{\alpha_m}] 
= i (-1)^{n+1} (4\pi)^{D/2} \int \frac{
    d^{D} P }{ (2\pi)^{D} }
  \frac{P^{\alpha_1} \cdots P^{\alpha_m}} {(P^2 - m^2) \cdots (
    (P-\sum_{i=1}^{n-1} k_i )^2 - m^2) } \,,
  \equn\label{GeneralLoop}
$$
where $P$ is a $D$-dimensional momentum and the external momenta,
$k_i$, are four-dimensional.  Our convention is to suppress the
dimension label on $I_n$ when we are dealing with a $D=4-2\eps$
integral, unless otherwise indicated.  We also suppress the square
brackets when the argument is unity. Such integrals are referred to in
the text as scalar integrals, while tensor integrals have $m\ge 1$.

Given a four-point amplitude, we have two conventions for labeling the
one-, two- and three-point integrals. The first, and more familiar one
is to explicitly give the kinematic invariant upon which the integral
depends. For example, $I_2(s)$ is the $(4-2\eps)$-dimensional bubble
that has an invariant mass square of $s=(k_1+k_2)^2$ flowing through
its external legs.  The second convention for labeling the lower point
integrals is to indicate, by a raised $(i)$ that the internal
propagator between legs $i-1$ (mod $4$) and $i$ has been removed.  If
two indices are present as in $(i,j)$ then two propagators have been
removed.  This follows the labeling convention of
refs.~\cite{IntegralRecursion,Massive}.

\subsection{The $D=4-2\eps$ scalar integrals}
\label{ScalarIntegralFunctions}

The bubble with an external kinematic invariant $s$ is 
$$
I_2(s) = I_2^{(2,4)} = I_2(0) 
+ 2 + x \log \left( {x-1 \over x+1} \right)
+ \Ord(\eps)\,,
\equn\label{EvalI2mass}
$$
where $x \equiv \sqrt{1 - 4m^2/s}$. 
For external massless kinematics, this may be written in closed form as
$$
I_2(0) = m^{-2\eps} \, { \Gamma(1+\eps) \over \eps } \, .
\equn\label{BubbleFour}
$$
The scalar triangle integral can either be a function of just 
one kinematic invariant $s$
$$
I_3(s) = 
I_3^{(4)} =
- {1\over 2s} \log^2 \left( {x+1 \over x-1} \right)
+ \Ord(\eps)\,,
\equn
$$
or of two invariants $s$ and $m_H^2$
$$
I_3(s,m_H^2) = I_3^{(2)} = \frac{sI_3(s)-m_H^2I_3(m_H^2)}{s-m_H^2}
+ \Ord(\eps)\,.
\equn\label{integral2invariants}
$$
While the box with a uniform internal mass and a single massive external leg, with invariant mass-square $m_H^2$ is
$$
I_4(s,t,m_H^2) = -{1\over st} \left[
     H \Bigl( -{u m^2\over st},{m^2\over s} \Bigr)
  +  H \Bigl(-{u m^2\over st},{m^2\over t}  \Bigr)
  -  H \Bigl(-{u m^2\over st},{m^2\over m_H^2}\Bigr)
\right] + \Ord(\eps) \,,
\equn\label{ExtMassBoxDef}
$$
where $u = m_H^2 - s - t$,  
$$
\eqalign{
H(X,Y) \equiv {2 \over x_+ - x_-} \biggl[
\ln \Bigl(1 - { X \over Y }\Bigr) \ln \Bigl({ -x_- \over x_+ }\Bigr)
&- {\rm Li}_2\Bigl({x_- \over y - x_+}\Bigr)
- {\rm Li}_2 \Bigl({x_- \over x_- - y}\Bigr)
\cr
&+ {\rm Li}_2\Bigl({x_+ \over x_+ - y}\Bigr)
+ {\rm Li}_2\Bigl({x_+ \over y - x_-}\Bigr)
\biggr] \, , 
}
\equn
$$
with $x_\pm = \half ( 1 \pm \sqrt{1-4X} )$, $y = \half ( 1 +
\sqrt{1-4Y} )$ and the dilogarithm \cite{Lewin} is defined by ${\rm
Li}_2(x) \equiv -\int_0^1 dt\; \ln(1-xt)/t$.

The scalar integral expressions quoted above are only valid in certain
analytic regions. These can be analytically continued, but are not
presented here since related expressions\footnote{The relationship
between the scalar integrals in this paper and those in
ref.~\cite{Ellis}: $I_2(s)=I_2(0)+2-W_1(s)$, $I_3(s)=-W_2(s)/2s$ and
$I_4(s,t,m_H^2)=-W_3(s,u,t,m_H^2)/st$.} are presented for all analytic
regions in ref.~\cite{Ellis}.

\subsection{Higher dimension integrals}
\label{HigherDimensionAppendix}

The higher dimension integrals are defined in \eqn{GeneralLoop} with
$D$ set to the appropriate value.  The relationship of these integrals
to the usual four-dimensional ones has been extensively discussed in
ref.~\cite{IntegralRecursion}.  For $n\le 6$, $(6-2\epsilon)$-dimensional
integrals can be written in terms of $(4-2\eps)$-dimensional integrals 
via the integral recursion relation
$$
\eqalign{
  I_n^{D=6-2\eps} & = {1\over (n-5+2\eps) c_0} \biggl[
         2 I_n^{D=4-2\eps}
 - \sum_{i=1}^n c_i I_{n-1}^{(i),D=4-2\eps}
  \biggr] \, , 
}
\equn
\label{HigherToLower}
$$ 
where  
$
  c_i = \sum_{j=1}^{4} S_{ij}^{-1} 
$
, 
$
c_0 = \sum_{i=1}^{n} c_i \, , 
$
and the matrix $S_{ij}$ is
$$
 S_{ij} \equiv  m^2 - {1 \over 2}  p_{ij}^2\,,  \hskip .3 cm 
 {\rm with} \hskip .3 cm 
   p_{ii}\ \equiv\ 0\,, \hskip .3  cm {\rm and} \hskip .3 cm 
  p_{ij} = p_{ji} \equiv
  k_i+k_{i+1}+\cdots+k_{j-1}\quad{\rm for}\ i<j \, .
\equn\label{SDef}
$$

Higher dimension integrals arise naturally when performing the 
calculations described in this paper. In a typical primitive amplitude  
discussed in the text we obtain integrals
of the form
$$
\eqalign{
  I_n^{D=4-2\epsilon}[&f(p^\alpha, k_i^\alpha,\mu^2)] =\cr
 & i (-1)^{n+1} (4\pi)^{2-\epsilon}
  \int \frac{ d^{4}p }{(2\pi)^{4} }
  \frac{d^{-2\epsilon}\mu}{(2\pi)^{-2\epsilon}}
  \frac{f(p^\alpha, k_i^\alpha, \mu^2)}{(p^2 - \mu^2 - m^2) \cdots
          ((p-\sum_{i=1}^{n-1} k_i )^2 - \mu^2 - m^2) } \, ,
}
\equn\label{BetterLoop}
$$
where we have explicitly broken the $(4-2\eps)$-dimensional momentum
into a four-dimensional part, $p$, and a $(-2\eps)$-dimensional part,
$\mu$.

As noted by Mahlon \cite{Mahlon}, using the definition
(\ref{GeneralLoop}), we can write
$$
I_n^{D=4-2\epsilon}[\mu^2] = -\epsilon I_n^{D=6-2\epsilon} \, .
\equn\label{DSEIntegrand}
$$
Note that although the loop momentum is shifted to higher 
dimension, the external momenta remain in four dimensions.

\subsection{Feynman parameter reduction}
\label{FeynmanParamAppendix}

As discussed in the text, tensor integrals may be evaluated using
Feynman parameters. We make use of the following formulae in the text.

After Feynman parameterization and the normal shift of momentum we
make use of the following,
$$
\eqalign{
\int {d^4 q\over (2\pi)^4}
 {d^{-2\epsilon}\mu \over (2\pi)^{-2\epsilon}}  \; 
 \frac{q^{\alpha_1} \cdots q^{\alpha_{2r+1}}}{(q^2 - \mu^2
- S_{ij} a_i a_j)^n} &=
  0 \, ,\cr}
\equn
$$
where $S_{ij}$ is defined in \eqn{SDef}. 

After integrating out the loop momenta we obtain integrals of the form
$$
I_n^D[f(a_k)] \equiv  \Gamma(n-D/2)\int_0^1 d^na_k\
   \delta (1 - {\textstyle \sum_r} a_r)
     { f(a_k) \over \left[ \sum_{i,j=1}^n S_{ij} a_i a_j - i\varepsilon
            \right]^{n-D/2}}\, .
\equn
$$
An integral reduction formula that is quite useful is
\cite{IntegralRecursion}
$$
  I_n^D[a_i]\ =\ {1\over2} \sum_{j=1}^n c_{ij}\ I_{n-1}^{(j), \, D}
   \ +\ {c_i\over c_0}\ I^D_n \, ,
\equn\label{LinearIntegralRelation}
$$
where
$c_{ij}\ =\ S^{-1}_{ij} - {c_ic_j\over c_0}\, , $ 
and $c_i$ and $c_0$ are defined in \app{HigherDimensionAppendix}.
Another useful integral reduction formula for $I_n^D[a_ia_j]$ can also
be found in \cite{IntegralRecursion}.

\section{Rules for $\mu$'s}
\label{MuSlashEtc}

To begin, we have a $(4-2\epsilon)$-dimensional vector, $Q^\alpha$,
which we can express as a sum of a four dimensional piece $q^\alpha$
and a $(-2\epsilon)$-dimensional $\mu^{\alpha}$; $Q^\alpha = q^\alpha
+ \mu^\alpha$. (We have used the convention that upper-case momenta
are $(4-2\epsilon)$-dimensional and lower-case momenta are
four-dimensional.) Here we summarize the rules for dealing with
$\mu$'s given in ref.~\cite{Massive}.

We abide by the conventional Dirac algebra, $\left\{\gamma^\alpha
,\gamma^\beta \right\}=2\eta^{\alpha\beta}$ and
$\gamma^{\alpha\dagger} \gamma^0 = \gamma^0 \gamma^{\alpha}$ , where
$\alpha$ and $\beta$ are $(4-2\epsilon$)-dimensional Lorentz
indices. The metric is given by $\eta^{\alpha\beta} = {\rm diag}
(+,-,-,-,\ldots)$. It follows that $Q\cdot Q = q^2 - \mu^2$ (the
negative sign preceding the $\mu^2$ is from the metric).

Thus $\s\mu$ anti-commutes with four-dimensional $\gamma^\alpha$
matrices, $\left\{ \s{q} , \s\mu \right\} = 0$.  So if we use the
conventions of 't~Hooft and Veltman \cite{DimensionalRegularization},
and adopt the arbitrary dimension definition, $ \gamma_5 = i \gamma^0
\gamma^1 \gamma^2 \gamma^3$, then $\left\{ \s{q} , \gamma_5 \right\} =
0$ and $\left[ \s{\mu} , \gamma_5 \right] = 0$.  That is $\s\mu$
freely commutes with $\gamma_5$, since $\gamma_5$ is constructed as a
product of four-dimensional Dirac matrices.  Hence, $\omega_\pm
\,\s{q} = \s{q} \,\omega_\mp$ and $\omega_\pm \, \s{\mu} = \s{\mu} \,
\omega_\pm$, where $\omega_\pm \equiv \frac{1}{2} ( 1 \pm\gamma_5 )$
is the helicity projection operator.  In calculations it is important
to first commute the $\s\mu$'s together before evaluating the
remaining four-dimensional pieces.

In the calculation presented in \sec{Calculation} the momentum of the
sewn fermion lines is $(4-2\epsilon)$-dimensional.  We will borrow the
bra and ket symbols to represent the spinors, but as helicity is not a
good quantum number we shall not label them with a helicity.  Since we
always sum over all states across the cut, there is no need to define
a $(4-2\epsilon)$ helicity notion.

In sewing the $(4-2\epsilon)$-spinors together we implicitly sum 
over the spin degrees of freedom, that is
$$
  \ket{Q} \bra{Q} = \s{Q} + m = \s{q} +\s{\mu} + m \, , 
  \mbox{~~~and~~~}
  \ket{-Q} \bra{-Q} = -\s{Q} + m = -\s{q} -\s{\mu} + m \, .
\equn\label{SpinSum}
$$
This notation glosses over the distinction of spinors and
anti-spinors. However, as discussed in ref.~\cite{Massive}, this may
introduce an overall sign which can be readily fixed by hand.
Alternatively, one may easily use both particle and anti-particle
spinors to obtain identical results.

\end{document}